\documentclass[onecolumn,amsmath,amssymb,nofootinbib,12pt]{article}
\usepackage{jheppub}
\usepackage{ifpdf}
\usepackage{graphicx,subcaption}
\usepackage{amsfonts}
\usepackage{amsmath}
\usepackage{amssymb}
\usepackage{epsfig}
\usepackage{pdfpages}
\usepackage{graphicx,epstopdf}
\usepackage[makeroom]{cancel}
\hypersetup{pdftex,colorlinks=true,allcolors=blue}
\usepackage{array}
\usepackage{ulem}

\newcommand{\fig}[1]{Fig.\ref{#1}}
\newcommand{\bra}[1]{|#1\rangle}
\def\be{\begin{equation}}
\def\ee{\end{equation}}
\def\ba{\begin{eqnarray}}
\def\ea{\end{eqnarray}}

\def\nn{\nonumber}
\def\lf{\left}
\def\rt{\right}

\newcommand{\eqs}[1]{Eqs. (\ref{#1})}
\newcommand{\eq}[1]{(\ref{#1})}

\def\nn{\nonumber}\def\lf{\left}\def\rt{\right}\def\q{\theta} \def\w{\omega}\def\r {\rho}  \def\y {\psi}   \def\p {\pi} \def\a {\alpha}  \def\d {\delta} \def\f {\phi} \def\g {\gamma} \def\h {\eta}  \def\k {\kappa} \def\l {\lambda}    \def\b {\beta}   \def\pd {\partial} \def \inf {\infty}  
\def\Q{\Theta} \def\W{\Omega}     \def\S {\Sigma} \def\D {\Delta}   \def\L {\Lambda}    \def\grad{\nabla}\def\.{\cdot}
\def\math {\mathcal}
\usepackage{xspace}
\begin{document}
\title{\Large Holographic complexity in charged Vaidya black hole}
\author[a]{Jie Jiang,}
\affiliation[a]{Department of Physics, Beijing Normal University,
Beijing 100875, China}
\emailAdd{jiejiang@mail.bnu.edu.cn}
\date{\today}
\abstract{In this paper, we use the ``complexity equals action" (CA) conjecture to discuss growth rate of the complexity in a charged AdS-Vaidya black hole formed by collapsing an uncharged spherically symmetric thin shell of null fluid. Using the approach proposed by Lehner $et\, al.$, we evaluate the action growth rate and the slope of the complexity of formation. Then, we demonstrate that the behaviors of them are in agreement with the switchback effect for the light shock wave case. Moreover, we show that to obtain an expected property of the complexity, it is also necessary for the CA conjecture to add the particular counterterm on the null boundaries.}
\maketitle

\section{Introduction}

In recent years, there has been a growing interest in the topic of ``quantum complexity" which is defined as the minimum number of gates required to obtain a target state starting from a reference state \cite{1,2}. In the holographic viewpoint, Brown $et\,al.$ suggested that the quantum complexity of the state in the boundary theory corresponds to some bulk gravitational quantities which are called ``holographic complexity". Then, the two conjectures: ``complexity equals volume" (CV) \cite{1,3} and ``complexity equals action" (CA) \cite{BrL,BrD}, were proposed. These conjectures have attracted many researchers to investigate the properties of both holographic complexity and circuit complexity in quantum field theory, $e.g.$, \cite{D,D1,D2,D3,D4,D5,D6,D7,D8,D9,D10,D11,D12,D13,D14,D15,D16,D17,D18,D19,D20,D21,D22,D23,D24,D25,D26,Jiang1,Jiang2,Jiang3,
Guo:2018kzl,Fan:2018wnv,Yang2018,An:2018dbz,Yang:2018nda,An:2018xhv,Fan:2018xwf,Feng:2018sqm}.

We only focus on the CA conjecture, which states that the complexity of a particular state $|\y(t_L,t_R)\rangle$ on the AdS boundary is given by
\ba
\math{C}\lf(|\y(t_L,t_R)\rangle\rt)\equiv\frac{S}{\p\hbar}\,,
\ea
where $S$ is the on-shell action in the corresponding Wheeler-DeWitt (WDW) patch, which is enclosed by the past and future light sheets sent into the bulk spacetime from the timeslices $t_L$ and $t_R$. In particular, it was found that there is a bound of the complexity growth rate at the late time
\ba
\dot{\math{C}}\leq \frac{2 M}{\p \hbar}\,,
\ea
which may be thought of as the Lloyd's bound on the quantum complexity \cite{E7}. As presented previously, at late times, the rate of the complexity will saturate this bound. However, by the full-time analysis \cite{D7}, we can see that this late time limit is approached from above, which will violate this bound.

In Refs.\cite{F1,F2}, Chapman $et\, al.$ investigated the CA and CV conjectures for AdS-Vaidya spacetime which is sourced by the collapse of a spherically symmetric thin shell of null fluid \cite{G1,G2,G3}. They found that the standard definition of the WDW action is not appropriate for these dynamical spacetimes. In order to obtain an expected property of the complexity, we need to add a particular counterterm on the null boundaries. This counterterm also keeps the invariance under the reparametrization of the null generator on the null boundary. Moreover, they also demonstrated that the switchback effect for light shocks are imprinted in the complexity of formation and the full-time evolution of complexity when this counterterm is introduced.

In this paper, we follow the discussions in \cite{F1,F2} to investigate the holographic complexity for a charged AdS-Vaidya black hole which is sourced by an uncharged thin shell. This thin shell will generate a shape transition from a black hole with total mass $M_1$ and  charge $Q$ to another one with mass $M_2$ and same charge $Q$. With the approach proposed by Lehner $et\, al.$ \cite{A5}, we will evaluate the time evolution of complexity growth rate as well as the slope of the complexity of formation in the presence of the light and heavy shock wave. Using these results, we will argue that our results are also in agreement with the switchback effect for the light shock wave case.

The structure of this paper is as follows. In Sec.\ref{sec2}, we review the charged AdS-Vaidya background geometries. In Sec.\ref{sec3}, we first use the method proposed by Lehner $et\, al.$ to calculate the complexity of formation as well as the action growth rate of the charged AdS-Vaidya black hole. Then, we investigate the action growth rate without the counterterm and compare our holographic results to the circuit behaviors. Concluding remarks are given in Sec.\ref{sec6}.

\section{Charged AdS-Vaidya spacetime}\label{sec2}
In this paper, we consider the $(d+1)$-dimensional Einstein-Maxwell gravity. Following the convention in Refs.\cite{A1,A2}, the total action can be expressed as
\ba\label{Sfull}
S_\text{total}=S_\text{grav}+S_\text{E.M.}+S_\text{ct}+S_\text{fluid}\,.
\ea
Here, the first two terms are the Einstein-Maxwell action which can be written as
\ba\label{fa}\begin{aligned}
&S_\text{grav}+S_\text{E.M.}=\frac{1}{16\p G}\int_{\math{M}} d^{d+1}x\sqrt{-g}\lf[(R-2\L)-F_{ab}F^{ab}\rt]+\frac{1}{8\p G}\int_{\math{B}} d^dx\sqrt{|h|} K\\
&+\frac{1}{8\p G}\int_{\S}d^{d-1}x\sqrt{\g}\h+\frac{1}{8\p G}\int_{\math{B}'} d\l d^{d-1}\q\sqrt{\g}\k+\frac{1}{8\p G}\int_{\S'}d^{d-1}x\sqrt{\g}a\,,
\end{aligned}\ea
where this action includes not only the bulk action of the Einstein-Maxwell theory but the surface terms and corner terms as well. The third term is the counterterm for the null boundaries \cite{D4}. It can be expressed as
\ba
S_\text{ct}=\frac{1}{8\p G}\int_{\math{B}'} d\l d^{d-1}\q\sqrt{\g}\Q\log\lf(l_{ct}\Q\rt)\,,
\ea
where $\Q=\pd_\l\ln\sqrt{\g}$ is the expansion scalar of the null surface generator, and $l_\text{ct}$ is an arbitrary constant length scale. This counterterm is added to keep the invariance under the reparametrization of the null generator.

The last term in (\ref{Sfull}) is the null fluid action. In order to construct an uncharged null fluid collapse, following the discussion in \cite{F1}, we can build the action by
\ba
S_{_\text{fluid}}=\int d^{d+1}x\sqrt{-g}\lf(\l g_{ab}l^al^b+sl^a\grad_a\f\rt)\,
\ea
with some real tensor fields.
According to the bulk action in (\ref{Sfull}), the equations of the motion can be expressed as
\ba\label{eom}\begin{aligned}
G_{ab}-\frac{2}{L^2}g_{ab}&=2 G \lf(F_{ac}F_b{}^{c}-\frac{1}{4}F^2g_{ab}\rt)+8\p G  T_{ab}\,,\\
\grad_aF^{ab}&=0\,,
\end{aligned}\ea
with
$T_{ab}=2\l l_a l_b$
which is the on-shell stress tensor of the null fluid. One solution is the charged AdS-Vaidya spacetime whose line element is given by
\ba\label{dsv}
ds^2=-F(r,v)dv^2+2dr dv +r^2d\S^2_{k,d-1}\,
\ea
with the blackening factor
\ba
F(r,v)&=&k+\frac{r^2}{L^2}-\frac{f_p(v)}{r^{d-2}}+\frac{q^2}{r^{2(d-2)}}\,.
\ea
Moreover, the corresponding Maxwell field and null fluid can be described by
\ba\begin{aligned}
A_a&=\sqrt{\frac{d-1}{2(d-2)}}\lf(\frac{q}{r_h^{d-2}}
-\frac{q}{r^{d-2}}\rt)(dv)_a\,,\\
\l&=\frac{(d-1)}{64\p G}\frac{ f_p'(v)}{r^{d-1}}\,,\\
l_a&=(dv)_a\,.
\end{aligned}\ea
where $r_h$ is the radius of the outer horizon.
This solution describes a spacetime which is sourced by the collapse of an uncharged spherically symmetric shell of null fluid. In particular, when the width of the shell shrinks to zero, this scalar function can be written as
\ba\label{Mv}
f_p(v)=w_1^{d-2}\lf[1-\math{H}(v-v_s)\rt]+w_2^{d-2} \math{H}(v-v_s)\,,
\ea
where $\math{H}(v)$ is the Heaviside step function. This function describes an infinitely thin shell collapse which generates a shape transition from a black hole with total mass $M_1$ and charge $Q$ to another one with mass $M_2$ and same charge $Q$, in which
\ba\begin{aligned}
M_i&=\frac{(d-1)\W_{k,d-1}}{16\p G}\w_i^{d-2}\,,\\
 Q&=\frac{\sqrt{2(d-1)(d-2)}\W_{k,d-1}}{8\p G}q\,,
\end{aligned}\ea
where $\W_{k,d-1}$ denotes the volume of the corresponding spatial geometry.

\section{Holographic complexity in charged AdS-Vaidya black hole}\label{sec3}
In this section, we turn to investigate the ``complexity equals action" conjecture. Following the standard procedures, we focus on the change rate of the action in the WDW patch of the charged AdS-Vaidya black hole with an uncharged thin shell of null fluid collapse. In this case, the WDW patch can be divided into three regions: the stationary region before the collapse, the null shell with a finite width, and the stationary region after the collapse. As shown in \cite{F1}, with the width of the shell shrinking to zero,  the contributions from the null shell will vanish. Thus, the full action only depends on other two stationary regions. According to the line element (\ref{dsv}), the on-shell bulk action can be expressed as
\ba
S_\text{bulk}=\frac{1}{16\p G}\int_V d^{d+1}x \sqrt{-g}\lf(-\frac{2 d}{L^2}+\frac{2(d-2)}{r^{2(d-1)}}q^2\rt)\,.
\ea
When the width of this shell shrinks to zero, the bulk metric can be described by (\ref{dsv}) with (\ref{Mv}). Then, the spacetime is divided into two regions by the null shell $v=v_s$. And the blackening factor can be written as
\ba\label{Fr}
v&<&v_s:  F(r,v)=f_1(r)=k+\frac{r^2}{L^2}-\frac{\w_1^{d-2}}{r^{d-2}}+\frac{q^2}{r^{2(d-2)}}\,,\nn\\
v&>&v_s: F(r,v)=f_2(r)=k+\frac{r^2}{L^2}-\frac{\w_2^{d-2}}{r^{d-2}}+\frac{q^2}{r^{2(d-2)}}\,.\nn\\
\ea
For the convenience of later calculations, we would like to introduce the tortoise coordinates as
\ba
v&<&v_s:\ \ \ \ \ r_1^*(r)=-\int_r^\inf \frac{dr}{f_1(r)}\,,\\
v&>&v_s:\ \ \ \ \ r_2^*(r)=-\int_r^\inf \frac{dr}{f_2(r)}\,.
\ea
We choose this range of integration to make that both expressions satisfy $\lim_{r\to\inf}r^*_{1,2}(r)\to 0$. According to Ref.\cite{D7}, with the blackening factors \eq{Fr}, one can obtain
\ba\begin{aligned}\label{rstar}
r^*_i(r)&=\frac{\ln(|r-r_{+,i}|/r)}{g_i(r_{+,i})(r_{+,i}-r_{-,i})}-\frac{\ln(|r-r_{-,i}|/r)}{g_i(r_{-,i})(r_{+,i}-r_{-,i})}-\frac{1}{r_{+,i}-r_{-,i}}\int^{\inf}_{r}G_i(r)dr\,,
\end{aligned}\ea
where
\ba\begin{aligned}
g_i(r)&=\frac{f_i(r)}{(r-r_{+,i})(r-r_{-,i})}\,,\\
G_i(r)&=\frac{g_i(r_{+,i})r-g_i(r)r_{+,i}}{g_i(r_{+,i})g_i(r)r(r-r_{+,i})}-\frac{g_i(r_{-,i})r-g_i(r)r_{-,i}}{g_i(r_{-,i})g_i(r)r(r-r_{-,i})}
\end{aligned}\ea
with $i=1,2$. Using these coordinates, one can also define an ``outgoing" null coordinate $u$ and auxiliary time coordinate $t$ as
\ba
u_{i}\equiv v-2r^*_i(r),\ \ \ \ \ \ \ t_i\equiv v-r^*_i(r)\,.
\ea
Next, we apply these coordinates to label the null surface which crosses the null shell at the point $r=r_w$. In the region $v>-t_w$, this surface can be described by $u_2=\bar{u}_2$. And in $v<-t_w$, it becomes $u_1=\bar{u}_1$. Since all of them cross the same point $(-t_w,r_w)$, we have
\ba
\bar{u}_2&=&-t_w-2r^*_2(r_w)\,,\label{uud}\\
\bar{u}_1&=&-t_w-2r^*_1(r_w)\,.\label{uud1}
\ea
By performing an infinitesimal transformation $\bar{u}_1\to \bar{u}_1+\d \bar{u}_1$ to this null surface and using (\ref{uud}) and (\ref{uud1}), one can further obtain
\ba\label{ts}
f_2(r_w) \d \bar{u}_2=f_1(r_w) \d \bar{u}_1\,.
\ea
Then, we introduce four positions which are important in defining the WDW patch in our case. As shown in Figs. (\ref{WdW1}) and (\ref{WdW2}), $r_b$ is where the left future boundary of the WDW patch meets the shock wave inside the future black hole, $r_s$ is where the right past boundary of the WDW patch meets the shock wave out of the black hole; $r_{1,2}$ is where the past/future null boundary segments of the WDW patch meet inside the horizon. In order to regulate the divergence near the AdS boundary, a cut-off surface $r=r_{\L}$ is introduced.

By using the tortoise coordinates, one can find that the coordinates $r_s,r_b,r_1$ and $r_2$ yield
\ba\begin{aligned}\label{rrrr}
t_w+2r_2^*(r_s)&=-t_R\,,\\
t_w+2r_1^*(r_b)&=t_L\,,\\
t_w+2r_1^*(r_s)&=t_L+2r_1^*(r_1)\,,\\
t_w+2r_2^*(r_b)&=-t_R+2r_2^*(r_2)\,.\\
\end{aligned}\ea

In what follows, we will use the methods in \cite{A5} to evaluate the derivative of the complexity of formation with respect to $t_w$ as well as the growth rate of the complexity in the charged Vaidya spacetime.

First of all, we consider the additional complexity, commonly referred to the complexity of formation, comes from the comparison of two circuit complexities, one is from thermofield double state (TFD), the other is two unentangled copies of the vacuum state, i.e.,
\ba
\D\math{C}=\math{C}(\bra{TFD})-\math{C}(\bra{0}_L\otimes\bra{0}_R)\,.
\ea
Using the CA conjecture, the holographic calculation is to evaluate the WDW action for $t_L=t_R=0$ in the black hole and subtract that for two copies of the AdS vacuum geometry. Note that the complexity of the formation can be studied as a function of $t_w$. In order to show the switchback effect, next, we consider the derivative of the complexity of formation with respect to $t_w$ (the slop of the complexity of formation). Through the shift symmetry to the antisymmetric time evolution of the complexity, we have
\ba\label{DDStw}
\frac{d\D S}{d t_w}=\left[\frac{dS}{d t_R}-\frac{dS}{d t_L}\right]_{t_L=t_R=0}\,,
\ea
where $S=S(t_L,t_R)$ is denoted as the action for the WDW patch determined by the time slices on the left and right AdS boundaries\cite{BrL,BrD}. Thus, the key to evaluate the slope of the complexity of formation is to obtain the time derivative of the action with respect to $t_R$ and $t_L$.

Then, we consider the growth rate of the complexity with respect to a symmetric time $t_L=t_R=t/2$. To evaluate this quantity, we turn attention to the change of the action which can be defined as $$\d S\equiv S(t_L+\d t/2,t_R+\d t/2)-S(t_L,t_R)\,$$ in the WDW patch. Following the standard prescription proposed by Refs.\cite{F1,F2}, we shall apply the affine parameter for null generator of null segments. As a consequence, the contributions from the corners at $r_{s/b}$, as well as all of the null segments will vanish. For simplicity, we rewrite the change of the action as $\d S=\d S_L+\d S_R$, with
\ba
\d S_L&=S(t_L+\d t/2,t_R)-S(t_L,t_R)\,,\\
\d S_R&=S(t_L,t_R+\d t/2)-S(t_L,t_R)\,.
\ea
Therefore, in order to obtain the slope of the complexity of formation as well as the growth rate of the complexity, we need derive the change of the action $\d S_R$ and $\d S_L$.
\subsection{The change of the action}
\subsubsection{$\d S_R$}\label{sec31}
\begin{figure}
\centering
\begin{minipage}{0.02\textwidth}
  \ \ \\
  \end{minipage}
\includegraphics[width=0.5\textwidth]{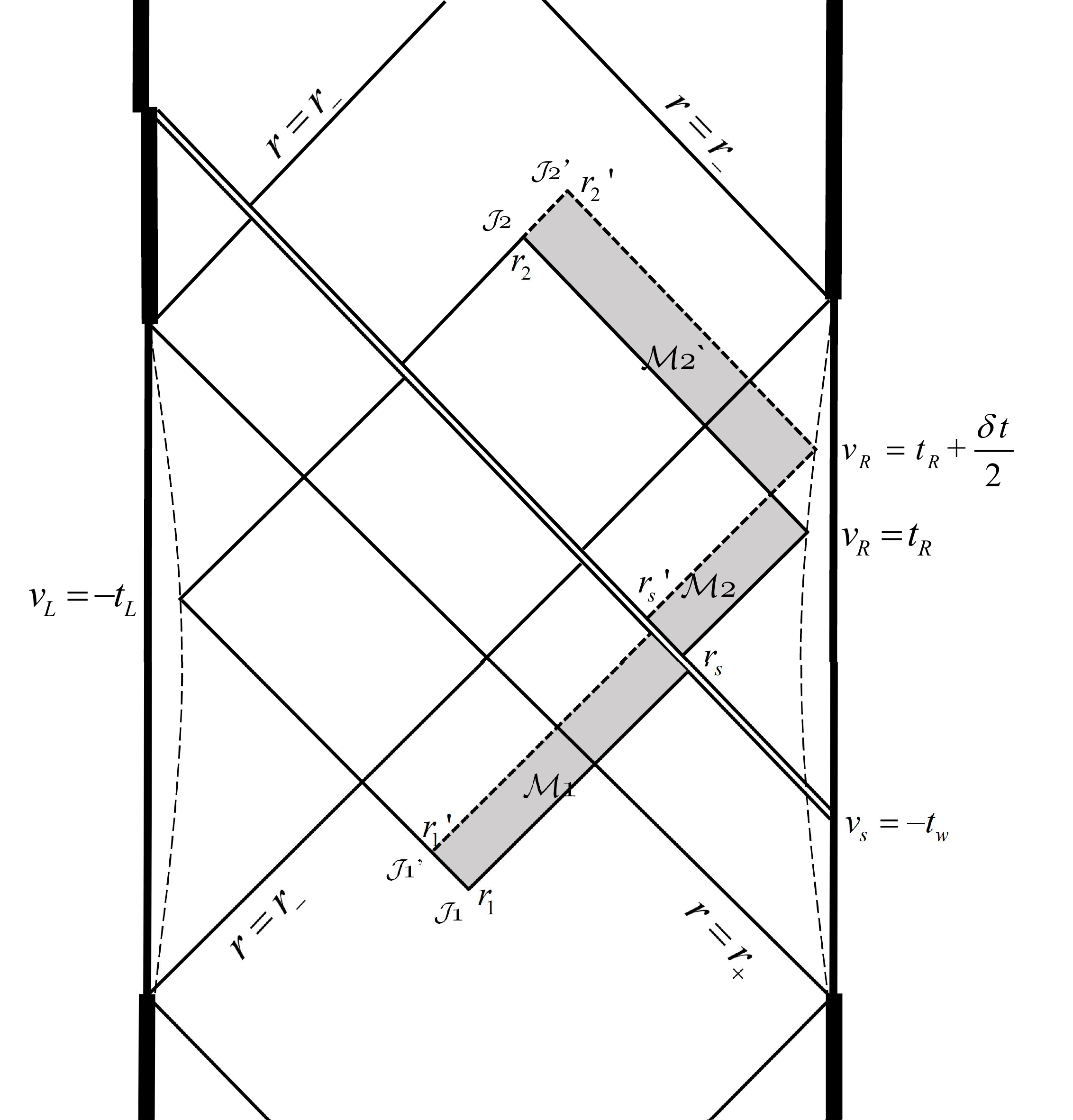}
\caption{The change of the Wheeler-DeWitt patches in a charged Vaidya-AdS black hole, where we fix the left boundary time $t_L$ and vary $t_R$ in the right boundary.}\label{WdW1}
\end{figure}
We first calculate $\d S_{R}$ where we fix the left boundary time $t_L$ and vary $t_R$ in the right boundary as shown in \fig{WdW1}. Considering the local symmetries of this spacetime, the nonvanish terms are contributed by the regions $\math{M}_1$, $\math{M}_2$, $\math{M}_2'$, as well as the joints $\math{J}_{1}'$, $\math{J}_{1}$, $\math{J}_{2}'$, $\math{J}_{2}$. Then, we have
\ba\begin{aligned}
\d S_R=S_{\math{M}_2'}-S_{\math{M}_2}+S_{\math{J}_2'}-S_{\math{J}_2}-S_{\math{M}_1}+S_{\math{J}_1'}-S_{\math{J}_1}\,.\nn\\
\end{aligned}\ea
Here,   $\math{M}_2$ is bounded by the null surfaces $v=t_R$, $v=-t_w$, $u_2=t_R$ and $u_2=t_R+\d t_R$. $\math{M}_2'$ is bounded by $u_2=t_R+\d t_R,u_2=u_L$, $v_2=t_R$ and $v_2=t_R+\d t_R$. And $\math{M}_1$ is bounded by $v=-t_L$, $v=-t_w$, $u_1=\bar{u}_1$ and $u_1=\bar{u}_1+\d \bar{u}_1$ with
\ba\begin{aligned}
\d \bar{u}_1=\frac{f_2(r_s)}{f_1(r_s)}\d \bar{u}_2=\frac{f_2(r_s)}{f_1(r_s)}\d t_R\,,
\end{aligned}\ea
where we have used \eq{ts}.

To evaluate the action contributed by $\math{M}_2'$, using the coordinate $(v,r)$ and keeping the first order of $\d t_R$, one can obtain
\ba\label{M22}\begin{aligned}
S_{\math{M}_2'}&=\frac{\W_{k,d-1}}{16\p G}\int_{t_R}^{t_R+\d t_R}dv \int_{\r_2'(v)}^{\r_2(v)}dr r^{d-1}\lf(-\frac{2 d}{L^2}+\frac{2(d-2)}{r^{2(d-1)}}q^2\rt)\\
&=-\frac{\W_{k,d-1}\d t_R}{8\p G}\lf(\frac{r_\L^d}{L^2}-\frac{r_{2}^d}{L^2}+ \frac{q^2}{r_\L^{d-2}}-\frac{q^2}{r_{2}^{d-2}}\rt)\,,
\end{aligned}\ea
where $r=\r_2(v)$ is the solution of the equation $u_2(v,r)=t_R+\d t_R$ and $r=\r_2'(r)$ is the solution of the equation $u_2(v,r)=u_L$. Similarly, with $(u_2,r)$ coordinates, we have
\ba\label{M2}\begin{aligned}
&S_{\math{M}_2}=-\frac{\W_{k,d-1}\d t_R}{8\p G}
\lf(\frac{r_\L^d}{L^2}-\frac{r_{s}^d}{L^2}+\frac{q^2}{r_\L^{d-2}}-\frac{q^2}{r_{s}^{d-2}}\rt)\,,
\end{aligned}\ea
where $r=\r(u_2),r=\r_s(u_2)$ are the solutions of the equation $v(u_2,r)=t_R$ and $v(u_2,r)=-t_w$, respectively. Let us
turn to the bulk region $\math{M}_1$. With similar calculation, one can further obtain
\ba\label{M1}\begin{aligned}
S_{\math{M}_1}=-\frac{\W_{k,d-1}}{8\p G}\d \bar{u}_1\lf(\frac{r_s^d}{L^2}-\frac{r_{1}^d}{L^2}+ \frac{q^2}{r_s^{d-2}}-\frac{q^2}{r_{1}^{d-2}}\rt)\,
\end{aligned}\ea
with $\d \bar{u}_1=\frac{f_2(r_s)}{f_1(r_s)}\d t_R$.
Combining these bulk contributions, we have
 \ba\begin{aligned}
 S_{\math{M}_2'}-S_{\math{M}_2}-S_{\math{M}_1}&=-\frac{\W_{k,d-1}\d t_R}{8\p G}\left\{\lf(1-\frac{f_2(r_s)}{f_1(r_s)}\rt)\frac{r_s^d}{L^2}-\frac{r_{2}^d}{L^2}+\frac{f_2(r_s)}{f_1(r_s)}\frac{r_{1}^d}{L^2}\right.\\
 &\left.+\lf[\lf(1-\frac{f_2(r_s)}{f_1(r_s)}\rt)\frac{1}{r_s^{d-2}}-\frac{1}{r_{2}^{d-2}}+\frac{f_2(r_s)}{f_1(r_s)}\frac{1}{r_{1}^{d-2}}\rt]q^2\right\}\,.\nn
\end{aligned}
\ea

We next consider the contributions from the joints in the $\d S_R$. Using the expression of the corner term, one can obtain
\ba
S_{\math{J}_{i}}= \frac{1}{8\p G} \int_{\math{J}_{i}}d^{d-1}x\sqrt{\g}\h_{i} =\frac{\W_{k,d-1}r_{i}^{d-1}}{8\p G}\h_{i}\,,
\ea
where  $\math{J}_{i}\in\{\math{J}_{1,2}, \math{J}_{1,2}'\}$ and $r_{i}\in\{r_{1,2}, r_{1,2}'\}$. To obtain the corner parameter $\h_i$, we need define the generator of the null boundary of WDW patch with affine parameters. The relevant null normals to the past right null boundary can be defined as
\begin{equation}\label{kp}
k_a^p = \left\{
             \begin{array}{lcl}
             \a \lf(-(dv)_a+\frac{2}{f_2(r)}(dr)_a\rt)\ \ \ \text{for} \ \ \ r>r_s \\
             \tilde{\a} \lf(-(dv)_a+\frac{2}{f_1(r)}(dr)_a\rt)\ \ \ \text{for} \ \ \ r<r_s
             \end{array}\,.
        \right.
\end{equation}
For the future left null boundary, we have
\begin{equation}\label{kf}
k_a^f = \left\{
             \begin{array}{lcl}
             \a \lf(-(dv)_a+\frac{2}{f_1(r)}(dr)_a\rt)\ \ \ \text{for} \ \ \ r>r_b \\
             \hat{\a} \lf(-(dv)_a+\frac{2}{f_2(r)}(dr)_a\rt)\ \ \ \text{for} \ \ \ r<r_b
             \end{array}\,.
        \right.
\end{equation}
By demanding that the null boundary is affinely parameterized across the shock wave, we have\cite{A1}
\ba
\frac{\tilde{\a}}{\a}=\frac{f_1(r_s)}{f_2(r_s)}\ \ \ \text{and}\ \ \ \ \frac{\hat{\a}}{\a}=\frac{f_2(r_b)}{f_1(r_b)}\,.
\ea
We can also introduce the null normal to the future right/past left null boundary,
\ba
k_a=\a (dv)_a\,.
\ea
In what follows, we consider the contributions from $\math{J}_2,\math{J}_2'$. Using $\h=\ln|\frac{1}{2}k_1\.k_2|$, one can obtain
\ba\begin{aligned}
\h_2'&= -\ln\lf(-\frac{f_2(r_2')f_1(r_b)}{\a^2f_2(r_b)}\rt)\,,\ \ \ \ \h_2= -\ln\lf(-\frac{f_2(r_2)f_1(r_b)}{\a^2f_2(r_b)}\rt)\,.
\end{aligned}\ea
Thus, we have
\ba\begin{aligned}
&S_{\math{J}_2'}-S_{\math{J}_2}=\frac{\W_{k,d-1}r_{2}'^{d-1}}{8\p G}\h_{{}_{2}}'-\frac{\W_{k,d-1}r_{2}^{d-1}}{8\p G}\h_{{}_{2}}\\
&=-\d t_R\lf[\frac{\W_{k,d-1}}{16\p G}r_{2}^{d-1} f_2'(r_{2})\right.\left.+\frac{\W_{k,d-1}(d-1)}{16\p G} \,r_{2}^{d-2} f_2(r_{2})\ln\lf(-\frac{f_2(r_{2})f_1(r_b)}{\a^2f_2(r_b)}\rt)\rt]\,,
\end{aligned}\ea
where we have used
\ba\label{dr2R}
\d r_2=r_2'-r_2=\frac{1}{2}f_2(r_2)\d t_R\,,
\ea
Then, we consider the contributions from $\math{J}_1,\math{J}_1'$. Using the relations
\ba\label{drsR}\begin{aligned}
\d r_s&=-\frac{f_2(r_s)}{2}\d t_R\,,\\
\d r_1&=-\frac{f_1(r_1)}{2}\d \bar{u}_1=-\frac{f_1(r_1)}{2}\frac{f_2(r_s)}{f_1(r_s)}\d t_R\,,
\end{aligned}\ea
and
\ba\begin{aligned}
\h_1'&= -\ln\lf(-\frac{f_1(r_1')f_2(r_s')}{\a^2f_1(r_s')}\rt)\,,\ \ \ \h_1 &= -\ln\lf(-\frac{f_1(r_1)f_2(r_s)}{\a^2f_1(r_s)}\rt)\,,
\end{aligned}\ea
one can obtain
\ba\begin{aligned}
&S_{\math{J}_1'}-S_{\math{J}_1}=\frac{\W_{k,d-1}r_{1}'^{d-1}}{8\p G}\h_{{}_{1}}'-\frac{\W_{k,d-1}r_{1}^{d-1}}{8\p G}\h_{{}_{1}}\\
&=\frac{\W_{k,d-1}\d t_R}{16\p G}\left[r_1^{d-1}\frac{f_2(r_s)f_1'(r_1)}{f_1(r_s)}\right.\left.+(d-1)r_1^{d-2}\frac{f_2(r_s)f_1(r_1)}{f_1(r_s)}\ln\lf(-\frac{f_1(r_1)f_2(r_s)}{\a^2 f_1(r_s)}\rt)\right]\\
&+\frac{\W_{k,d-1}\d t_Rr_1^{d-1}}{16\p G}\lf(f_2'(r_s)-f_1'(r_s)\frac{f_2(r_s)}{f_1(r_s)}\rt)\,.
\end{aligned}\ea
Combining these expressions, we have
\ba\label{dSR}\begin{aligned}
&\d S_R=S_{\math{M}_2'}-S_{\math{M}_2}-S_{\math{M}_1}+S_{\math{J}_2'}-S_{\math{J}_2}+S_{\math{J}_1'}-S_{\math{J}_1}\\
&=-\frac{\W_{k,d-1}}{16\p G}\left\{2\lf(1-\frac{f_2(r_s)}{f_1(r_s)}\rt)\lf(\frac{r_s^d}{L^2}+\frac{q^2}{r_s^{d-2}}\rt)\right.\left.-\lf(d-2\rt)\lf(\w_2^{d-2}-\frac{f_2(r_s)}{f_1(r_s)}\w_1^{d-2}\rt)\rt\}\d t_R\\
&-\frac{(d-1)\W_{k,d-1}}{8\p G}\lf(\frac{f_2(r_s)}{f_1(r_s)}\frac{q^2}{r_1^{d-2}}-\frac{q^2}{r_2^{d-2}}\rt)\d t_R+\frac{\W_{k,d-1}r_1^{d-1}}{16\p G}\lf(f_2'(r_s)-f_1'(r_s)\frac{f_2(r_s)}{f_1(r_s)}\rt)\d t_R\\
&-\frac{(d-1)\W_{k,d-1}}{16\p G}\left[r_{2}^{d-2} f_2(r_{2})\ln\lf(-\frac{f_2(r_{2})f_1(r_b)}{\a^2f_2(r_b)}\rt)\right.\left.-r_1^{d-2}\frac{f_2(r_s)f_1(r_1)}{f_1(r_s)}\ln\lf(-\frac{f_1(r_1)f_2(r_s)}{\a^2 f_1(r_s)}\rt)\right]\d t_R
\end{aligned}\ea
\begin{figure}
\centering
\begin{minipage}{0.02\textwidth}
  \ \ \\
  \end{minipage}
\includegraphics[width=0.5\textwidth]{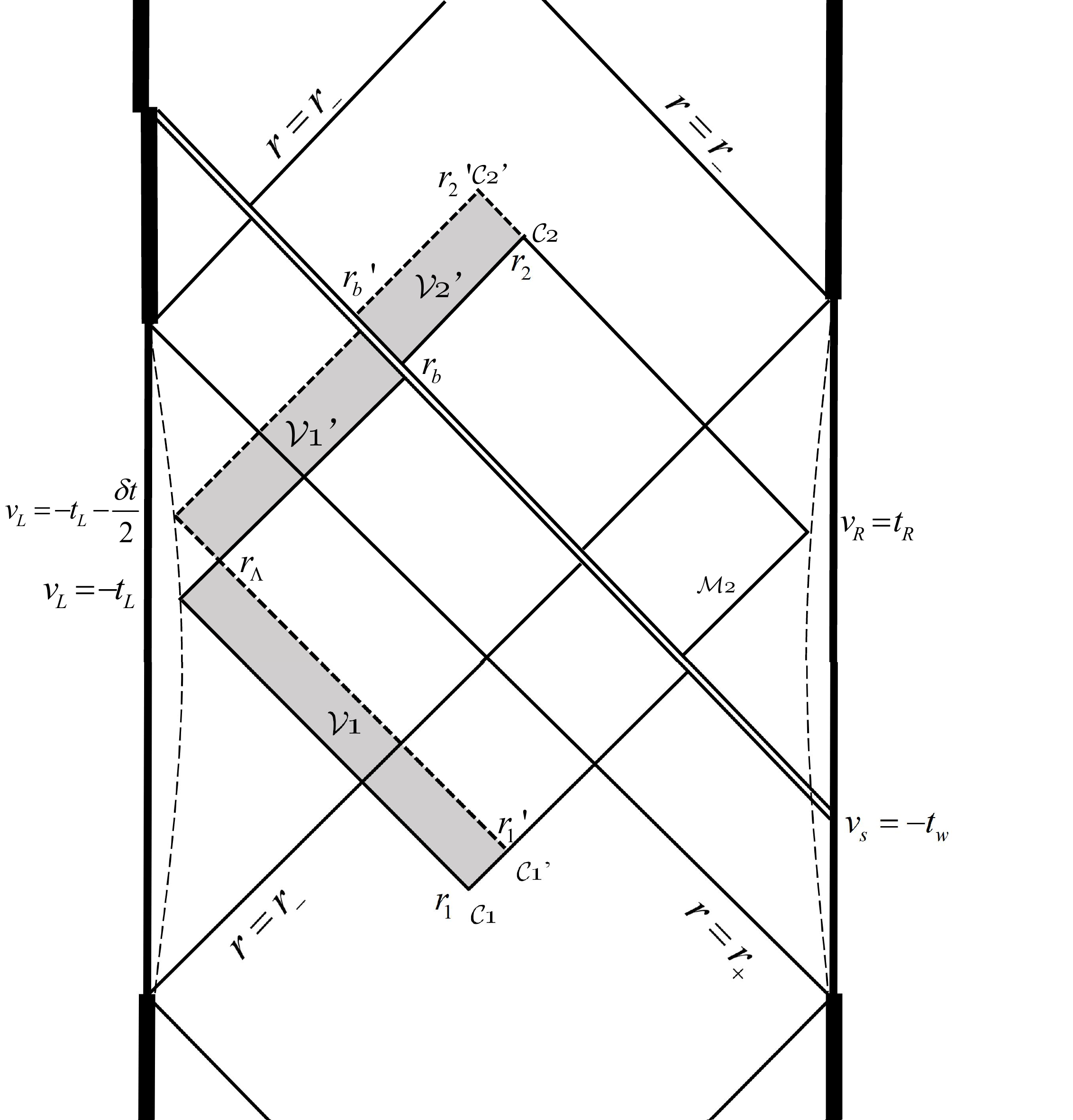}
\caption{The change of the Wheeler-DeWitt patches in a charged Vaidya-AdS black hole, where we fix the right boundary time $t_R$ and vary $t_L$ in the left boundary.}\label{WdW2}
\end{figure}
\subsubsection{$\d S_L$}\label{sec32}
We turn to calculate $\d S_{L}$ where we fix the right boundary time $t_R$ and vary $t_L$ in left boundary as illustrated in \fig{WdW2}. The nonvanish terms are contributed by the regions $\math{V}_1$, $\math{V}_2$, $\math{V}_2'$, as well as the joints $\math{C}_{1}'$, $\math{C}_{1}$, $\math{C}_{2}'$, $\math{C}_{2}$. Then, we have
\ba
\d S_L=S_{\math{V}_2'}+S_{\math{V}_1'}-S_{\math{V}_1}+S_{\math{C}_2'}-S_{\math{C}_2}+S_{\math{C}_1'}-S_{\math{C}_1}\,.
\ea
Turning to the bulk contributions, with similar calculation, one can obtain
\ba\label{V11}\begin{aligned}
S_{\math{V}_1'}&=-\frac{\W_{k,d-1}\d t_L}{8\p G}\lf(\frac{r_\L^d}{L^2}-\frac{r_b^d}{L^2}+\frac{q^2}{r_\L^{d-2}}-\frac{q^2}{r_b^{d-2}}\rt)\,,\\
S_{\math{V}_1}&=-\frac{\W_{k,d-1}\d t_L}{8\p G}\lf(\frac{r_\L^d}{L^2}-\frac{r_1^d}{L^2}+\frac{q^2}{r_\L^{d-2}}-\frac{q^2}{r_1^{d-2}}\rt)\,,\\
S_{\math{V}_2'}&=-\frac{\W_{k,d-1}\d t_L}{8\p G}\frac{f_1(r_b)}{f_2(r_b)}\lf(\frac{2r_b^d}{L^2}-\frac{2r_2^d}{L^2}+\frac{q^2}{r_b^{d-2}}-\frac{q^2}{r_2^{d-2}}\rt)\,.\nn\\
\end{aligned}\ea
Using the relations
\ba
\d r_b = \frac{f_1(r_b)}{2}\d t_L\,,\ \ \ \d r_2=\frac{f_2(r_2)}{2}\frac{f_1(r_b)}{f_2(r_b)}\d t_L\,,
\ea
the corner terms which are contributed by the joints $\math{C}_1'$ ,$\math{C}_1$, $\math{C}_2'$ and $\math{C}_2$ can be expressed as
\ba\begin{aligned}
&S_{\math{C}_1'}-S_{\math{C}_1}=\frac{\W_{k,d-1}\d t_L}{16\p G}\lf[r_1^{d-1}f_1'(r_1)+(d-1)r_1^{d-2}f_1(r_1)\ln\lf(-\frac{f_1(r_1)f_2(r_s)}{\a^2 f_1(r_s)}\rt)\rt]\\
&S_{\math{C}_2'}-S_{\math{C}_2}=-\frac{\W_{k,d-1}\d t_Lr_1^{d-1}}{16\p G}\lf(f_1'(r_b)-f_2'(r_b)\frac{f_1(r_b)}{f_2(r_b)}\rt)\\
&-\frac{\W_{k,d-1}\d t_L}{16\p G}\frac{f_1(r_b)}{f_2(r_b)}\left[r_2^{d-1}f_2'(r_2)+(d-1)r_2^{d-2}f_2(r_2)\ln\lf(-\frac{f_2(r_2)f_1(r_b)}{\a^2 f_2(r_b)}\rt)\right]\\
\end{aligned}
\ea
Combining these expressions, we have
\ba\label{dSL}\begin{aligned}
&\d S_L=\frac{\W_{k,d-1}}{16\p G}\left[2\lf(1-\frac{f_1(r_b)}{f_2(r_b)}\rt)\lf(\frac{r_b^d}{L^2}+\frac{q^2}{r_b^{d-2}}\rt)\right.\left.+\lf(d-2\rt)\lf(\w_1^{d-2}-\frac{f_1(r_b)}{f_2(r_b)}\w_2^{d-2}\rt)\rt]\d t_L\\
&+\frac{(d-1)\W_{k,d-1}}{8\p G}\lf(\frac{f_1(r_b)}{f_2(r_b)}\frac{q^2}{r_2^{d-2}}-\frac{q^2}{r_1^{d-2}}\rt)\d t_L-\frac{\W_{k,d-1}r_2^{d-1}}{16\p G}\lf(f_1'(r_b)-f_2'(r_b)\frac{f_1(r_b)}{f_2(r_b)}\rt)\d t_L\\
&+\frac{(d-1)\W_{k,d-1}}{16\p G}\left[r_1^{d-2} f_1(r_1)\ln\lf(-\frac{f_1(r_1)f_2(r_s)}{\a^2f_1(r_s)}\rt)\right.\left.-r_2^{d-2}\frac{f_1(r_b)f_2(r_2)}{f_2(r_b)}\ln\lf(-\frac{f_2(r_2)f_1(r_b)}{\a^2 f_2(r_b)}\rt)\right]\d t_L\,.
\end{aligned}\ea

\subsubsection{Counterterm contributions}\label{sec33}
In this subsection, we calculate the contributions from the counterterm as mentioned above. In our case, we need to consider the contributions from all of the null boundaries of the WDW patch. First, we consider the past null boundary on the right side of the WDW patch. As illustrated in \fig{WdW2}, this boundary crosses the shock wave at $r=r_s$. From \eq{kp}, the null normal of this null surface can be re-expressed by
\ba
k^p_a=H(r,v)\lf(-(dv)_a+\frac{2}{F(r,v)}(dr)_a\rt)
\ea
with affine parameters, where we denote
\ba
H(r,v)=\a \math{H}(r-r_s)+\tilde{\a}\lf(r-r_s\rt)\,.
\ea
Due to $k^a=\lf(\frac{\pd}{\pd \l}\rt)^a$, one can obtain $dr/d\l=H(r,v)$. Using the expression
$
\Q=k^a\grad_a \ln\sqrt{\g}\,,
$
the expansion scalar of this null surface generators can be further expressed by
\ba
\Q=\frac{(d-1)H(r,v)}{r}\,.
\ea
Whence, the counterterm contribution for the past null boundary on the right side can be written as
\ba\label{Sct1}\begin{aligned}
S_\text{ct}^{(1)}&=\frac{\W_{k,d-1}(d-1)}{8\p G}\int_{r_1}^{r_\L}dr\,r^{d-2}\ln\left(\frac{(d-1)l_{ct}H(r,v)}{r}\right)\\
&=\frac{\W_{k,d-1}}{8\p G}\lf[r_{\L}^{d-1}\ln\lf(\frac{(d-1)\a l_{ct}}{r_{\L}}\rt)\right.\left.-r_1^{d-1}\ln\lf(\frac{(d-1)\a l_{ct}}{r_1}\rt)+\frac{r_\L^{d-1}-r_1^{d-1}}{d-1}\rt]\\
&+\frac{\W_{k,d-1}}{8\p G}\lf(r_s^{d-1}-r_1^{d-1}\rt)\ln\lf(\frac{f_1(r_s)}{f_2(r_s)}\rt)\,.
\end{aligned}
\ea
where we replaced $d\l=dr/H(r,v)$. Next, we consider the left future boundary of the WDW patch. By replacing $r_s, r_1$ with $r_b,r_2$ respectively, the corresponding conterterm can be further obtained
\ba\label{Sct2}\begin{aligned}
S_\text{ct}^{(2)}
&=\frac{\W_{k,d-1}}{8\p G}\lf[r_{\L}^{d-1}\ln\lf(\frac{(d-1)\a l_{ct}}{r_{\L}}\rt)\right.\left.-r_2^{d-1}\ln\lf(\frac{(d-1)\a l_{ct}}{r_2}\rt)+\frac{r_\L^{d-1}-r_2^{d-1}}{d-1}\rt]\\
&+\frac{\W_{k,d-1}}{8\p G}\lf(r_b^{d-1}-r_2^{d-1}\rt)\ln\lf(\frac{f_2(r_b)}{f_1(r_b)}\rt)\,.
\end{aligned}\ea
With similar calculation, counterterm contributions of the past boundary on the left side and the future boundary on the right can be expressed as
\ba\label{Sct3}\begin{aligned}
&S_\text{ct}^{(3)}=\frac{\W_{k,d-1}}{8\p G}\lf[r_{\L}^{d-1}\ln\lf(\frac{(d-1)\a l_{ct}}{r_{\L}}\rt)\right.\left.-r_{1}^{d-1}\ln\lf(\frac{(d-1)\a l_{ct}}{r_{1}}\rt)+\frac{r_\L^{d-1}-r_1^{d-1}}{d-1}\rt]\,.\\
&S_\text{ct}^{(4)}=\frac{\W_{k,d-1}}{8\p G}\lf[r_{\L}^{d-1}\ln\lf(\frac{(d-1)\a l_{ct}}{r_{\L}}\rt)\right.\left.-r_{2}^{d-1}\ln\lf(\frac{(d-1)\a l_{ct}}{r_2}\rt)+\frac{r_\L^{d-1}-r_2^{d-1}}{d-1}\rt]\,.
\end{aligned}\ea

Then, we consider the change of the action where we fix the right boundary time $t_R$ and vary $t_L$ in the left boundary. Using \eqs{dr2R} and \eqs{drsR}, one can obtain
\ba\begin{aligned}\label{dSctR}
&\d S_\text{ctR}=\lf(r_1^{d-1}-r_s^{d-1}\rt)\frac{\W_{k,d-1}}{16\p G}\lf[\frac{f_2(r_s)f_1'(r_s)}{f_1(r_s)}-f_2'(r_s)\rt]\d t_R\\
&+\frac{(d-1)\W_{k,d-1}}{16\p G}\lf\{r_1^{d-2}f_1(r_1)\frac{f_2(r_s)}{f_1(r_s)}\ln\lf(\frac{f_1(r_s)}{f_2(r_s)}\rt)\right.+2r_1^{d-2}f_1(r_1)\frac{f_2(r_s)}{f_1(r_s)}\ln\lf(\frac{(d-1)\a l_{ct} }{r_1}\rt)\\
&-r_2^{d-2}f_2(r_2)\lf[\ln\lf(\frac{f_2(r_b)}{f_1(r_b)}\rt)+2\ln\lf(\frac{(d-1)\a l_{ct}}{r_2}\rt)\right]\left.-r_s^{d-2}f_2(r_s)\ln\lf(\frac{f_1(r_s)}{f_2(r_s)}\rt)\rt\}\d t_R\,.
\end{aligned}\nn\\\ea
When we fix the left boundary time $t_L$ and vary $t_R$, the corresponding change of the action can be shown as
\ba\begin{aligned}\label{dSctL}
&\d S_\text{ctL}=\frac{(1-d)\W_{k,d-1}}{16\p G}\lf\{r_2^{d-2}f_2(r_2)\frac{f_1(r_b)}{f_2(r_b)}\ln\lf(\frac{f_2(r_b)}{f_1(r_b)}\rt)\right.+2r_2^{d-2}f_2(r_2)\frac{f_1(r_b)}{f_2(r_b)}\ln\lf(\frac{(d-1)\a l_{ct} }{r_2}\rt)\\
&-r_1^{d-2}f_1(r_1)\lf[\ln\lf(\frac{f_1(r_s)}{f_2(r_s)}\rt)+2\ln\lf(\frac{(d-1)\a l_{ct}}{r_1}\rt)\right]\left.-r_b^{d-2}f_1(r_b)\ln\lf(\frac{f_2(r_b)}{f_1(r_b)}\rt)\rt\}\d t_L\\
&-\lf(r_2^{d-1}-r_b^{d-1}\rt)\frac{\W_{k,d-1}}{16\p G}\lf[\frac{f_1(r_b)f_2'(r_b)}{f_2(r_b)}-f_1'(r_b)\rt]\d t_L\,.
\end{aligned}\ea
\begin{figure*}
\centering
\includegraphics[width=3in,height=2in]{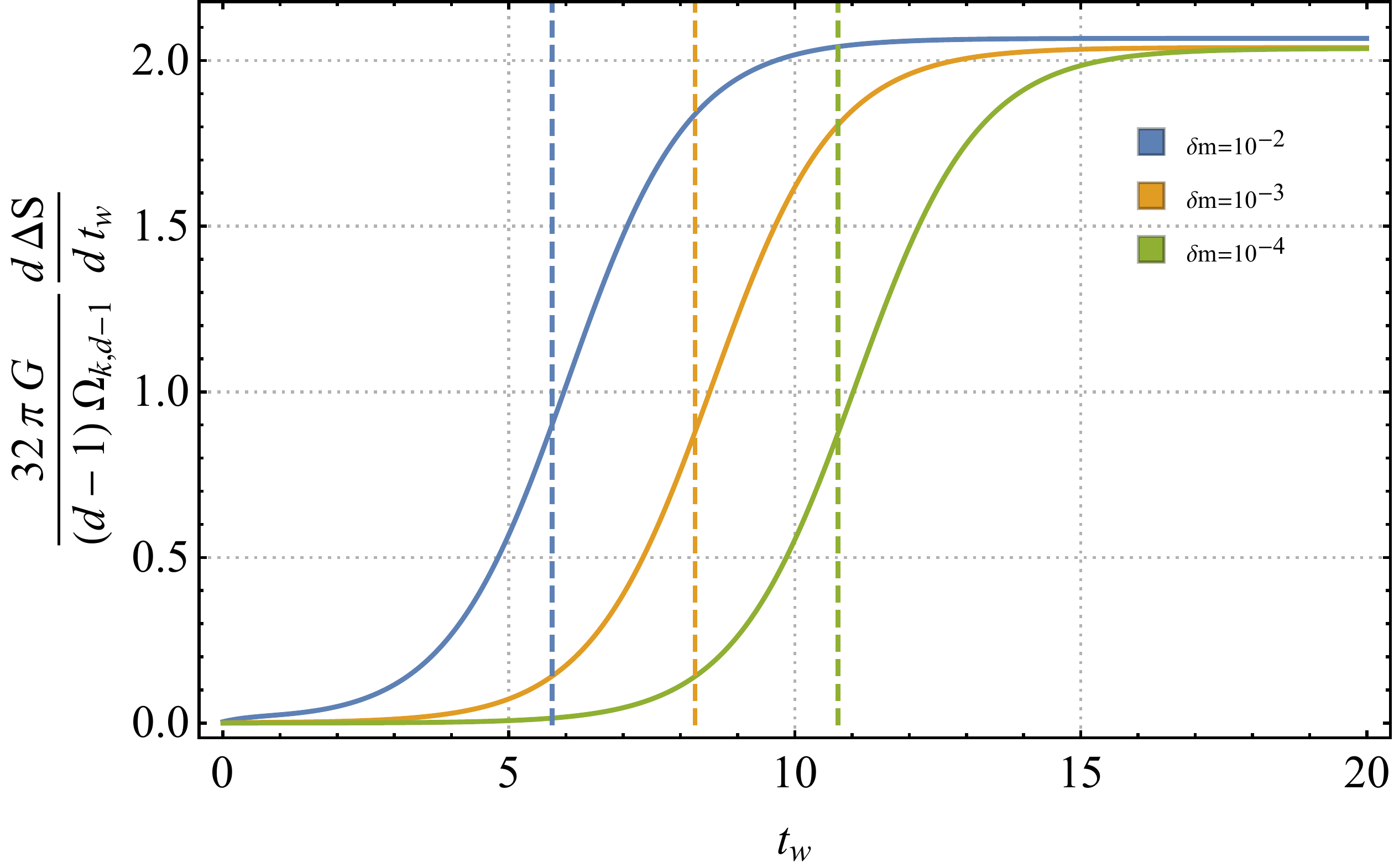}
\includegraphics[width=3in,height=2.05in]{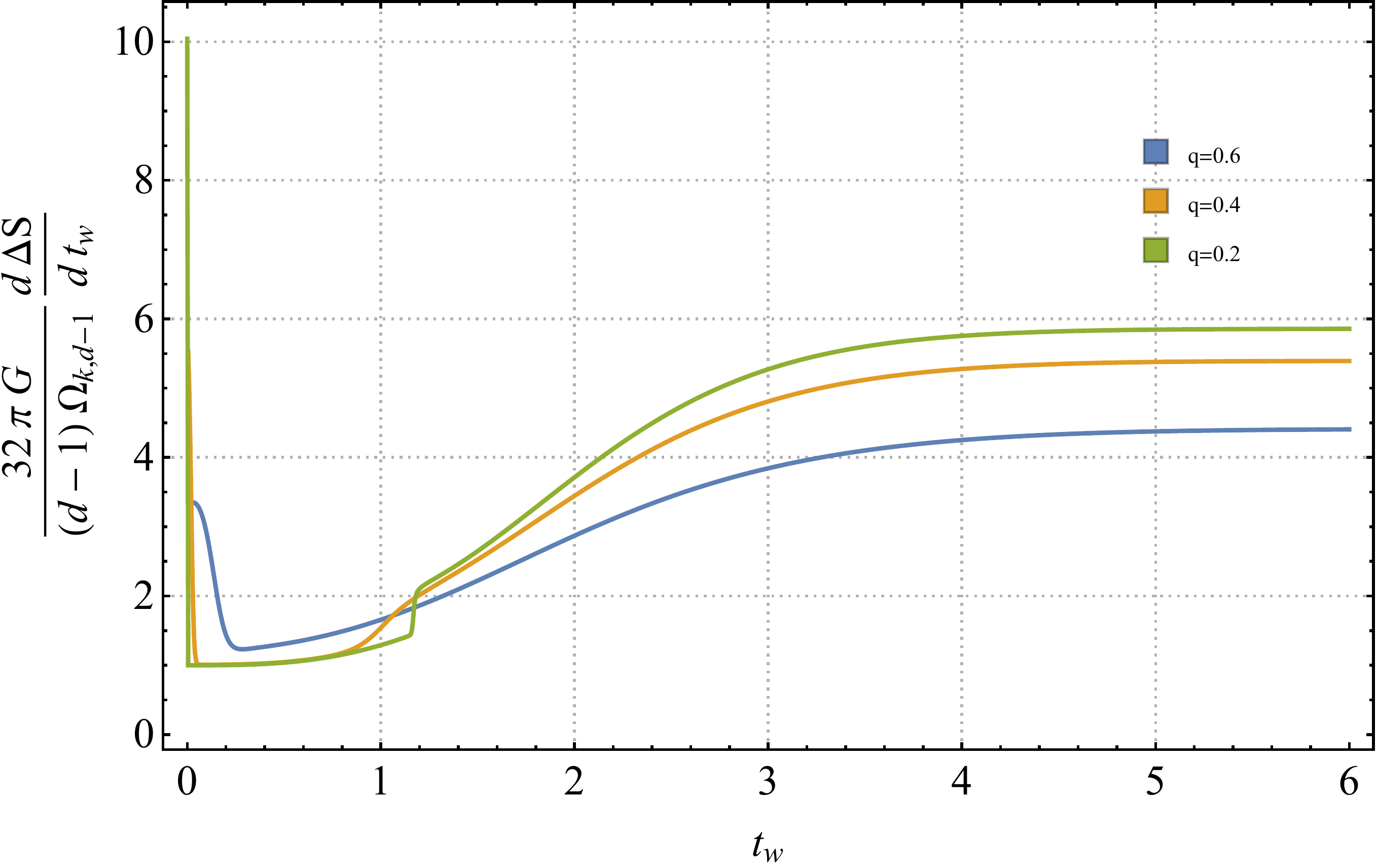}
\caption{The derivative of the complexity of formation with respect to $t_w$. The left panel illustrates the behaviour for a light shock wave, with $d=3, k=0, l_\text{ct}=1, L=1,q=0.6,\w_1=1$, where the dishes lines describe the corresponding scrambling time $t=t^*_\text{scr}$. The right panel illustrates the behaviour for the heavy shock wave with $d=3, k=0, l_\text{ct}=1, L=1,\w_1=1,w_2=2$}.
\label{dC}
\end{figure*}
\subsection{Complexity of Formation}\label{complexiyofformation}
In this subsection, we consider the complexity of formation. By using Eqs. \eqs{DDStw},\eqs{dSR},\eqs{dSL},\eqs{dSctR}, and \eqs{dSctL}, one can obtain
\ba\begin{aligned}
&\frac{32\p G}{{(d-1)}\W_{k,d-1}}\frac{d\D S}{d t_w}=2\lf(\frac{q^2}{r_2^{d-2}}+\frac{q^2}{r_1^{d-2}}-\frac{q^2}{r_b^{d-2}}-\frac{q^2}{r_s^{d-2}}\rt)-2\frac{f_1(r_b)}{f_2(r_b)}\lf(\frac{q^2}{r_2^{d-2}}-\frac{q^2}{r_b^{d-2}}\rt)\\
&-2\frac{f_2(r_s)}{f_1(r_s)}\lf(\frac{q^2}{r_1^{d-2}}-\frac{q^2}{r_s^{d-2}}\rt)+\lf[r_b^{d-2}f_1(r_b)\ln\lf(\frac{f_2(r_b)}{f_1(r_b)}\rt)-r_s^{d-2}f_2(r_s)\ln\lf(\frac{f_1(r_s)}{f_2(r_s)}\rt)\rt]\\
&+\lf[r_1^{d-2}f_1(r_1)\lf(\frac{f_2(r_s)}{f_1(r_s)}-1\rt)\ln\lf(-\frac{(d-1)^2l_{ct}^2f_1(r_1)}{r_1^2}\rt)\rt]\\
&-\lf[r_2^{d-2}f_2(r_2)\lf(1-\frac{f_1(r_b)}{f_2(r_b)}\rt)\ln\lf(-\frac{(d-1)^2l_{ct}^2f_2(r_2)}{r_2^2}\rt)\rt]\,.\nn
\end{aligned}\ea

Using Eqs. \eqs{rstar} and \eqs{rrrr}, the slope of the complexity of formation can be directly evaluated.
In the left panel of \fig{dC}, we show the effect of a light shock wave on the slope of the complexity of formation as a function of $t_w$. As shown in this figure, one can find that there exists a scrambling time $t_{\text{scr}}^*$ which is characterized by the energy of the shock wave $\d\w=\w_2-\w_1$. And the slope is approximately zero until the $t\simeq t_{\text{scr}}^*$ at which point it rapidly rises to the final constant value. This implies that for the order of the scrambling time $t_{\text{scr}}^*$, the complexity of formation is same as the case of unperturbed state. In the regime of $t_w>t_{\text{scr}}^*$, it grows linearly with respect to the time $t_w$. This shares the similar behavior with the uncharged black hole in \cite{F2}. And it is also in agreement with the switchback effect which we will discuss in Sec.\ref{circuit}.

In the right panel of \fig{dC}, we show the effect of heavier shock waves. In this regime, the slope starts at a finite value and suddenly drop to a minimal value, after that, it rapidly rises to the final constant value. It implies that the complexity of formation starts changing immediately and rapidly approach a regime of linear growth with increasing $t_w$. This is very different with the light shock wave case.

Now, we would like to analytically investigate the behaviour of these figures in the case of the light shock wave with $\frac{\w^{d-2}_2}{\w_1^{d-2}}=1+2\epsilon$. In order to find the scrambling time for the light shock wave, we consider the limit where the shock wave enters at very early time, i.e., $t_w\gg 1$. According to \eq{rstar}, one can obtain
\ba
r_s/r_+\approx 1+e^{-8\p T_1 t_w}\,,\ \ r_b/r_+\approx 1-e^{-8\p T_1 t_w}\,.
\ea
In this limit, there are two interesting regimes: $\epsilon\ll e^{-8\p T_1 t_w}$ and $\epsilon \gg e^{-8\p T_1 t_w}$. Then, the scrambling time $t_\text{scr}=-\frac{1}{8\p T_1}\ln \epsilon$ is determined by the transition condition $\epsilon\approx e^{-8\p T_1 t_w}$.

\subsubsection{Large and small time behaviors}
According to these figures, one can see that there exist two interesting regimes: $t_w\ll t_{\text{scr}}^*$ and $t_w\gg t_{\text{scr}}^*$, i.e., the small and large limit of $t_w$. First, we consider the small time limit. In this limit, we have $t_w\to0$, which will give $r_s\to \inf$ and $r_b,r_1,r_2\to r_m$. Then, we have
\ba\begin{aligned}
&\frac{32\p G}{{(d-1)}\W_{k,d-1}}\left.\frac{d\D S}{d t_w}\right|_{t_w\to0^+}=\w_1^{d-2}-\w_2^{d-2}\\
&-r_m^{d-2}f_2(r_m)\ln\lf(-\frac{(d-1)^2f_2(r_m)l_{ct}}{r_m^2}\rt)+r_m^{d-2}f_1(r_m)\ln\lf(-\frac{(d-1)^2f_1(r_m)l_{ct}}{r_m^2}\rt)\,,
\end{aligned}\ea
In the limit of the light shock wave, we have $\w_2\simeq \w_1$. Then, the slope will approach zero, which is in agreement with the behavior as illustrated in the left panel of \fig{dC}.

Next, we consider the large time limit $t_w\to \inf$. In this limit, $r_s$ and $r_b$ approach $r_{+,2}$ and $r_{+,1}$ respectively. With these in mind, we have
\ba\begin{aligned}
&\frac{32\p G}{{(d-1)}\W_{k,d-1}}\left.\frac{d\D S}{d t_w}\right|_{t_w\to\inf}=\left.2\lf(\frac{q^2}{r_2^{d-2}}+\frac{q^2}{r_1^{d-2}}-\frac{q^2}{r_{+,1}^{d-2}}-\frac{q^2}{r_{+,2}^{d-2}}\rt)\right.\\
&-r_1^{d-2}f_1(r_1)\ln\lf(-\frac{(d-1)^2l_{ct}^2f_1(r_1)}{r_1^2}\rt)\left.-r_2^{d-2}f_2(r_2)\ln\lf(-\frac{(d-1)^2l_{ct}^2f_2(r_2)}{r_2^2}\rt)\rt.\,,
\end{aligned}\ea
In limit of the light shock wave, by replacing the label $2$ to $1$, we can further obtain
\ba
\frac{32\p G}{{(d-1)}\W_{k,d-1}}\left.\frac{d\D S}{d t_w}\right|_{t_w\to\inf}=4\lf(\frac{q^2}{r_m^{d-2}}-\frac{q^2}{r_{+}^{d-2}}\rt)
-2r_m^{d-2}f(r_m)\ln\lf(-\frac{(d-1)^2l_{ct}^2f(r_m)}{r_m^2}\rt)\,.\nn\\
\ea
Here, we also used the relation $r_1=r_2=r_m$ under the light shock wave limit.

\begin{figure*}
\centering
\includegraphics[width=3in,height=2in]{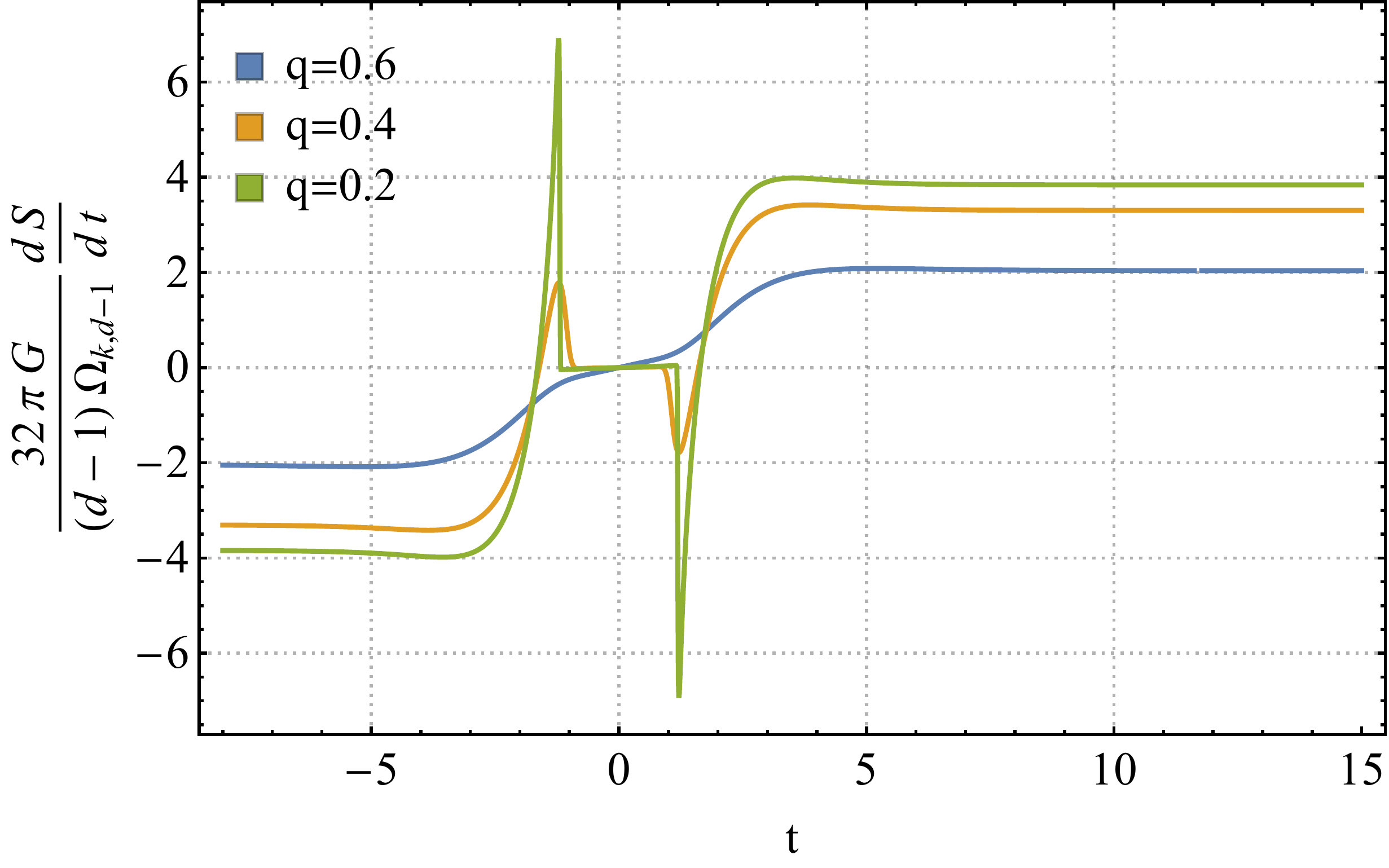}
\includegraphics[width=3in,height=2in]{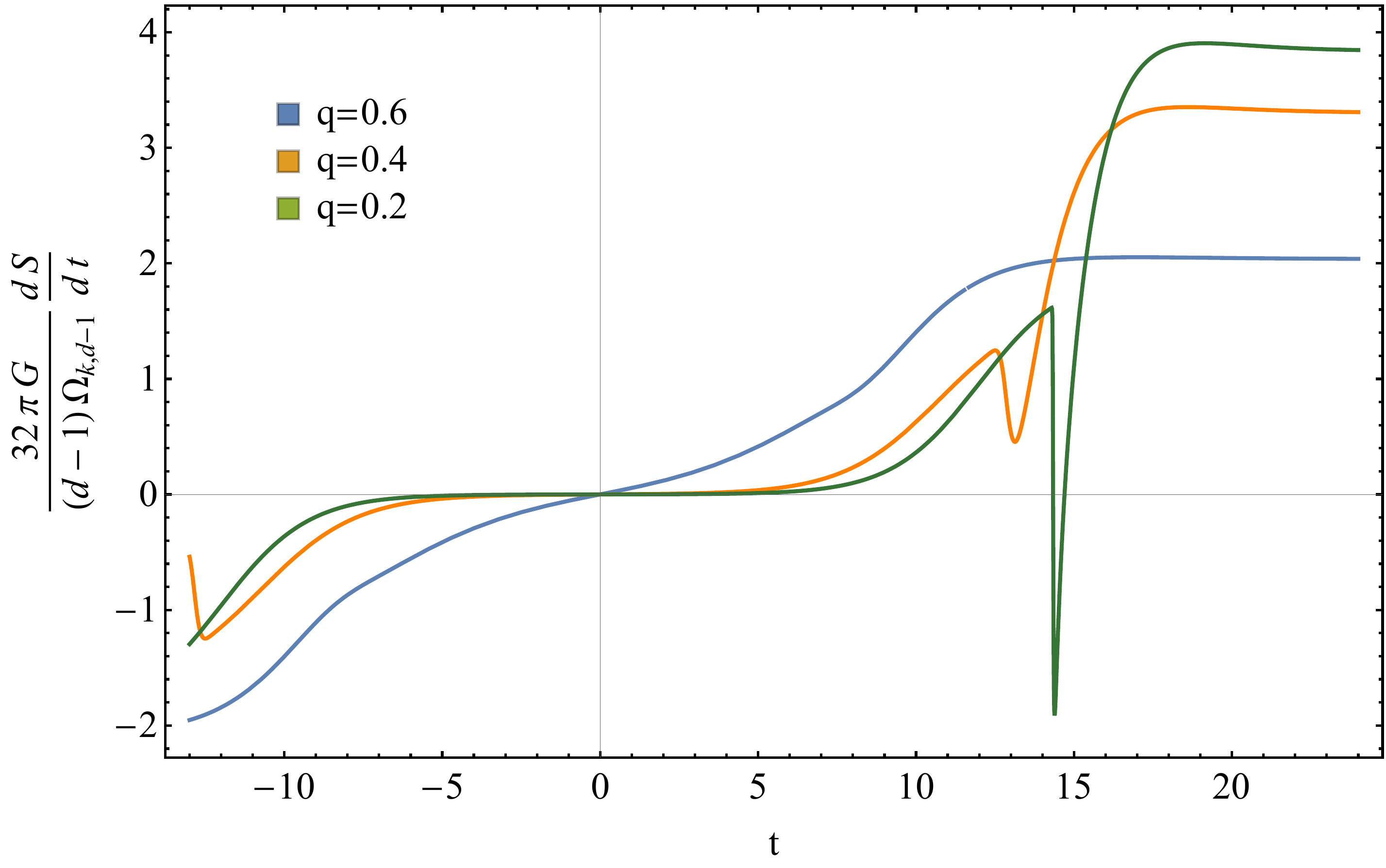}
\caption{The action growth rate for the light shock wave with the symmetric time evolution as $t_L=t_R=t/2$ with $t_w=5$ (left) and $t_w=14$ (right). We have set $d=3, k=0, l_\text{ct}=1, L=1,q=0.6,\w_1=1,\w_2=1+10^{-4}$}
\label{LdCdttw}
\end{figure*}

\subsection{Time evolution of the complexity}\label{evocom}
\begin{figure}
\centering
\includegraphics[width=3in,height=2.05in]{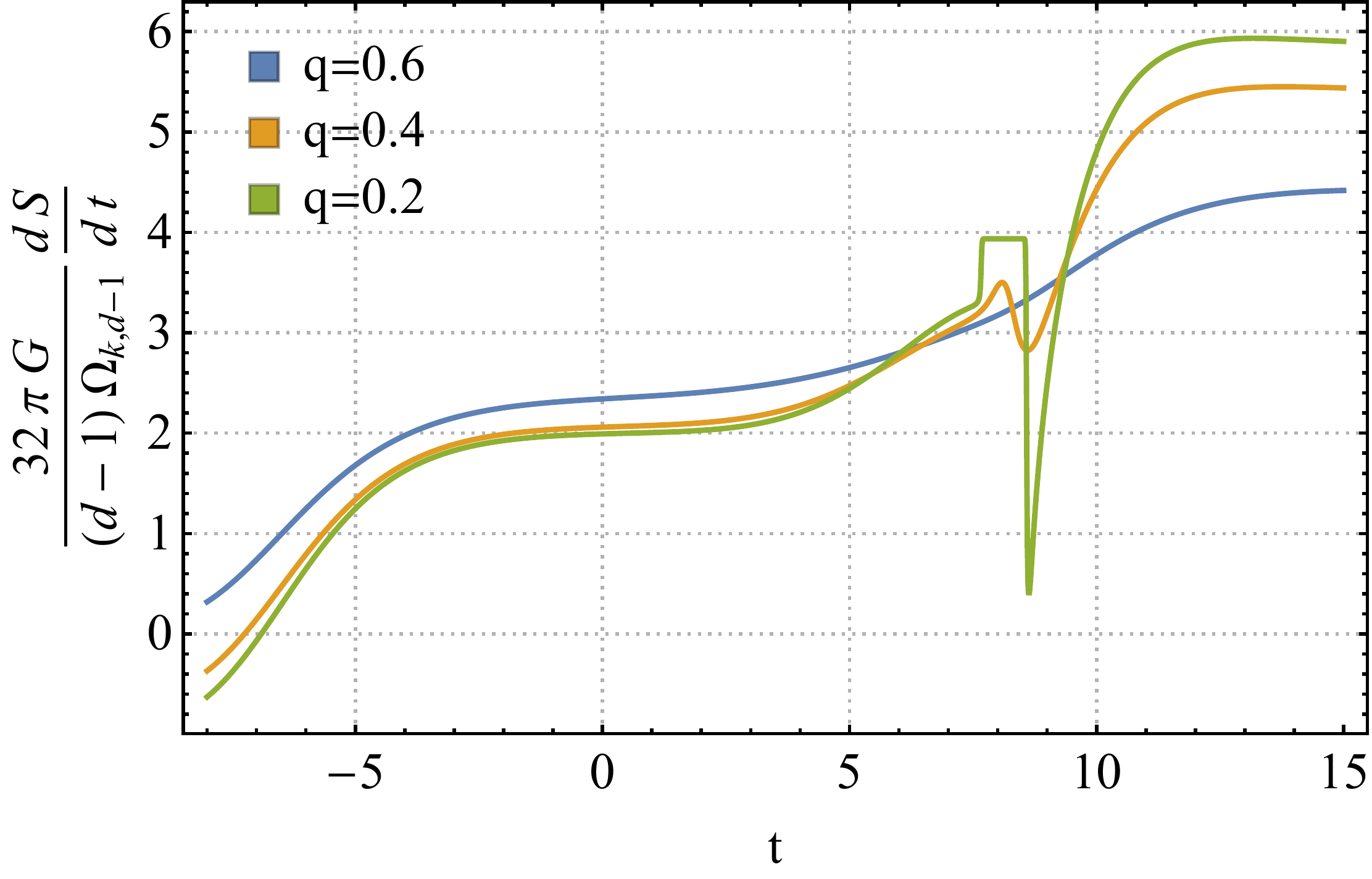}
\caption{The action growth rate for the heavy shock waves with the symmetric time evolution as $t_L=t_R=t/2$, where we set $d=3, k=0, l_\text{ct}=1, L=1,q=0.6,\w_1=1,\w_2=1+10^{-4}$, and $t_w=5$. Here $q=0.687$ is a special case where the initial black hole is a extremal black hole.}
\label{dCdttw}
\end{figure}
In this subsection, we consider the time evolution of the holographic complexity. By summing the various expressions above, the actiom growth rate with the counterterm can be written as
\ba\begin{aligned}\label{actiongrowthrate}
&\frac{32\p G}{{(d-1)}\W_{k,d-1}}\frac{d S}{d t}=2\lf(\frac{q^2}{r_2^{d-2}}-\frac{q^2}{r_1^{d-2}}+\frac{q^2}{r_b^{d-2}}-\frac{q^2}{r_s^{d-2}}\rt)
+2\frac{f_1(r_b)}{f_2(r_b)}\lf(\frac{q^2}{r_2^{d-2}}-\frac{q^2}{r_b^{d-2}}\rt)\\
&-2\frac{f_2(r_s)}{f_1(r_s)}\lf(\frac{q^2}{r_1^{d-2}}-\frac{q^2}{r_s^{d-2}}\rt)+\lf[r_b^{d-2}f_1(r_b)\ln\lf(\frac{f_2(r_b)}{f_1(r_b)}\rt)-r_s^{d-2}f_2(r_s)\ln\lf(\frac{f_1(r_s)}{f_2(r_s)}\rt)\rt]\\
&+\lf[r_1^{d-2}f_1(r_1)\lf(1+\frac{f_2(r_s)}{f_1(r_s)}\rt)\ln\lf(-\frac{(d-1)^2l_{ct}^2f_1(r_1)}{r_1^2}\rt)\rt]\\
&-\lf[r_2^{d-2}f_2(r_2)\lf(1+\frac{f_1(r_b)}{f_2(r_b)}\rt)\ln\lf(-\frac{(d-1)^2l_{ct}^2f_2(r_2)}{r_2^2}\rt)\rt]\,.\nn\\
\end{aligned}\ea
Under the limit of the light shock wave, the time dependent action growth rate will return to that of the eternal RN black hole \cite{D7}.

Considering Eqs. \eqs{rstar} and \eqs{rrrr}, we can numerically calculate the action growth rate in \eq{actiongrowthrate}. Then, we show the action growth rate as the function of $t$ for the light and heavy shock wave in \fig{LdCdttw} and \fig{dCdttw} separately. In these figures, we can see that the action growth rate develops a minimum or maximum at some finite time in very small charge case. These minimum or maximum becomes deeper and sharper for smaller charges. Therefore, the behaviors for the charged cases can smoothly approach that of the neutral cases. And the minimum or maximum is corresponding to the critical time in uncharged black hole \cite{F2}.

In the left panel of \fig{LdCdttw}, we show the action growth rate for a very light shock wave with $\d\w=\w_2-\w_1=10^{-4}$ at $t_w=5$. These figures show the same pictures with that of the internal RN black hole, which can be understood by the switch back effect since $t_w=5<t_\text{scr}^*$ in this case. In the right panel, we show the growth rate at $t_w=14$ such that $t_w>t^*_\text{scr}$. After the scrambling time, the action of the light shock wave will be clearly illustrated. Therefore, in this case, it will share the similar behaviours with the case of heavy shock wave as shown in \fig{dCdttw}. Moreover, for the small charge case, a minimum value of the action growth rate appears at a finite time. Under the uncharged limit, this minimum point will reduce to the critical time in the neutral case as shown in Fig.2 of Ref.\cite{F2}.

In \fig{dCdttw}, we show the action growth rate for a heavier shock wave with $\d \w =1$ at $t_w=5$. We can see that there might exist two critical times under the uncharged limit, which will coincide with the neutral case for the heavier shock wave in Fig.3 of Ref.\cite{F2}. In addition, as shown in \fig{dCdttw}, for the non-extremal case, there exists two horizontal periods, in which the rate can be regarded as constant. However, for the extremal case, there only exists one horizontal period, i.e., the late time period.

\subsubsection{Early and late time behaviors}
Here, we consider some simple limits for the growth rate of the complexity. First, we begin by examining the early time behavior, where $t_w$ is sufficiently large. Then $r_s$ approaches $r_{+,2}$ and $r_b$ approach $r_{+,1}$. Then, the growth rate of the complexity becomes
\ba\begin{aligned}
&\frac{32\p G}{{(d-1)}\W_{k,d-1}}\left.\frac{d S}{d t}\right|_{t_w\to \inf}=2\lf(\frac{q^2}{r_2^{d-2}}-\frac{q^2}{r_1^{d-2}}+\frac{q^2}{r_{+,1}^{d-2}}-\frac{q^2}{r_{+,2}^{d-2}}\rt)\\
&+r_1^{d-2}f_1(r_1)\ln\lf(-\frac{(d-1)^2l_{ct}^2f_1(r_1)}{r_1^2}\rt)-r_2^{d-2}f_2(r_2)\ln\lf(-\frac{(d-1)^2l_{ct}^2f_2(r_2)}{r_2^2}\rt)\,.
\end{aligned}\ea
One can find that this limit depends on the value of the times $t_R$ and $t_L$, which is different from the uncharged case where this limit is simply proportional to the difference of the masses.

Next, we consider the late time behaviors. In the late time limit, the points $r_b, r_s ,r_1$ and $r_2$ approach to $r_{-,1}$,$r_{+,2}$,$r_{+,1}$ and $r_{-,2}$ respectively. As a consequence, we have $f_1(r_b), f_1(r_1), f_2(r_s), f_2(r_2)\to 0 $. Using these expressions, the action growth rate can be written as
\ba\label{dcl}\begin{aligned}
&\frac{32\p G}{{(d-1)}\W_{k,d-1}}\left.\frac{dS}{dt}\right|_{t\to\inf}=2\lf(\frac{q^2}{r_{-,1}^{d-2}}-\frac{q^2}{r_{+,1}^{d-2}}
+\frac{q^2}{r_{-,2}^{d-2}}-\frac{q^2}{r_{+,2}^{d-2}}\rt)\,.
\end{aligned}\ea
The late time rate is proportional to the average value of the two eternal RN-AdS rate without shockwave with parameter $1$ and $2$. It would be convenient to work in terms of the following dimensionless quantities:
\ba\begin{aligned}
&y=\frac{r_{-,2}}{r_{+,2}}\,,\ \ \ \ \ \a=\frac{r_{+,2}}{r_{+,1}}\,, \ \ \ \ \ \b=\frac{r_{-,1}}{r_{-,2}}\,,\ \ \ \ \ z=\frac{L}{r_{+,2}}\,,\ \ \ \ \ x=\frac{r}{r_{+,1}}\,.
\end{aligned}\ea
Using the black hole mass and these dimensionless quantities, according to (\ref{dcl}), one can obtain
\ba
\left.\frac{dC_A}{dt}\right|_{t\to\inf}=\frac{\L_1M_1+\L_2M_2}{\p}\,
\ea
with
\ba
\L_1=\frac{\lf(1-y_1^{d-2}\rt)\lf[\lf(1-y_1^d\rt)+k z_1^2\lf(1-y_1^{d-2}\rt)\rt]}{\lf(1-y_1^{2(d-1)}\rt)+kz_1^2\lf(1-y_1^{2(d-2)}\rt)}\,,\\
\L_2=\frac{\lf(1-y^{d-2}\rt)\lf[\lf(1-y^d\rt)+k z^2\lf(1-y^{d-2}\rt)\rt]}{\lf(1-y^{2(d-1)}\rt)+kz^2\lf(1-y^{2(d-2)}\rt)}\,,
\ea
in which
\ba
y_1=\frac{r_{-,1}}{r_{+,1}}=\a\,\b\, y\,,\ \ \ \ \ z_1=\frac{L}{r_{+,1}}=\a\, z\,.
\ea
In these variables, when we set $\w_2\to \w_1$, i.e., $\a,\b\to1$, this result will return to the case with the light shock wave. Meanwhile, it is also equal to the value of the eternal RN black hole \cite{D7}. When we set $y\to 0$, this result will return to that of the uncharged case \cite{F2}. For the cases $k=0,1$, it's not difficult to see that our late time value is less than the uncharged case, i.e., this result saturates the bound
\ba
\left.\frac{dC_A}{dt}\right|_{t\to\inf}\leq\frac{M_1+M_2}{\p}\,.
\ea

\subsection{Complexity without counterterm}\label{without}
In this subsection, we consider the growth rate of the complexity where we drop the counterterm from the full action. Without the inclusion of the counterterm, the growth rate is only contributed by $\d S_R$ and $\d S_L$. Considering the late time limit, from Eqs. \eqs{dSL} and \eqs{dSR}, one can obtain
\ba\label{DS2}\begin{aligned}
&\left.\frac{d\tilde{S}}{dt}\right|_{t\to\inf}=\frac{\W_{k,d-1}}{32\p G}\left[2\lf(\frac{r_{-,1}^d-r_{+,2}^d}{L^2}+\frac{q^2}{r_{-,1}^{d-2}}-\frac{q^2}{r_{+,2}^{d-2}}\rt)\right.\\
&+r_{+,1}^{d-1}f'_2(r_{+,2})-r_{-,2}^{d-1}f'_1(r_{-,1})\left.+\lf(d-2\rt)\lf(\w_1^{d-2}-\w_2^{d-2}+\frac{2q^2}{r_{-,2}^{d-2}}-\frac{2q^2}{r_{+,1}^{d-2}}\rt)\right]\,.\\
\end{aligned}\ea
First, we consider the limit of light but still non-zero shocks. In this limit, we have $\w_2\simeq\w_1$, $r_{+,2}\simeq r_{+,1}$, $f_2\simeq f_1$ and $r_{-,2}\simeq r_{-,1}$. Then, the late time limit becomes
\be\label{DS2}
\frac{d\tilde{S}}{dt}=\frac{\W_{k,d-1}}{32\p G}\left[\frac{(3d-4)q^2}{r^{d-2}}-(d-2)\lf(k+\frac{r^2}{L^2}\rt)r^{d-2}\right]^{r_{-,1}}_{r_{+,1}}\,.\\
\ee
Next, we consider the shock wave with exactly zero energy. In this situation, we have $\w_2=\w_1$ and $f_2= f_1$. According to (\ref{dSL}) and (\ref{dSR}), the action growth rate can be shown as
\ba\begin{aligned}
&\frac{d\tilde{S}}{dt}=\frac{(d-1)\W_{k,d-1}}{8\p G}\lf(\frac{q^2}{r_2^{d-2}}-\frac{q^2}{r_1^{d-2}}\rt)\\
&+\frac{(d-1)\W_{k,d-1}}{16\p G}\left[r_1^{d-2}f_1(r_1)\ln\lf(-\frac{f_1(r_1)}{\a^2 }\rt)\right.\left.-r_{2}^{d-2} f_2(r_{2})\ln\lf(-\frac{f_2(r_{2})}{\a^2}\rt)\right]\,
\end{aligned}\ea
which is exactly the growth rate of the eternal RN black hole as discussed in \cite{D7}. Then, the late time limit can be given by
\ba\label{dds2}\begin{aligned}
\frac{d\tilde{S}}{dt}=&\frac{(d-1)\W_{k,d-1}}{8\p G}\lf(\frac{q^2}{r_{-,1}^{d-2}}-\frac{q^2}{r_{+,1}^{d-2}}\rt)\,.
\end{aligned}\ea
Comparing (\ref{DS2}) and (\ref{dds2}), one can find that the late time growth rate in the limit of light shocks can't return to the case without shock wave. Therefore, in order to obtain an expected property of the complexity, it is necessary to add the counterterm into the full action for the CA conjecture.

\begin{figure}
\centering
\begin{minipage}{0.02\textwidth}
  \ \ \\
  \end{minipage}
\includegraphics[width=0.5\textwidth]{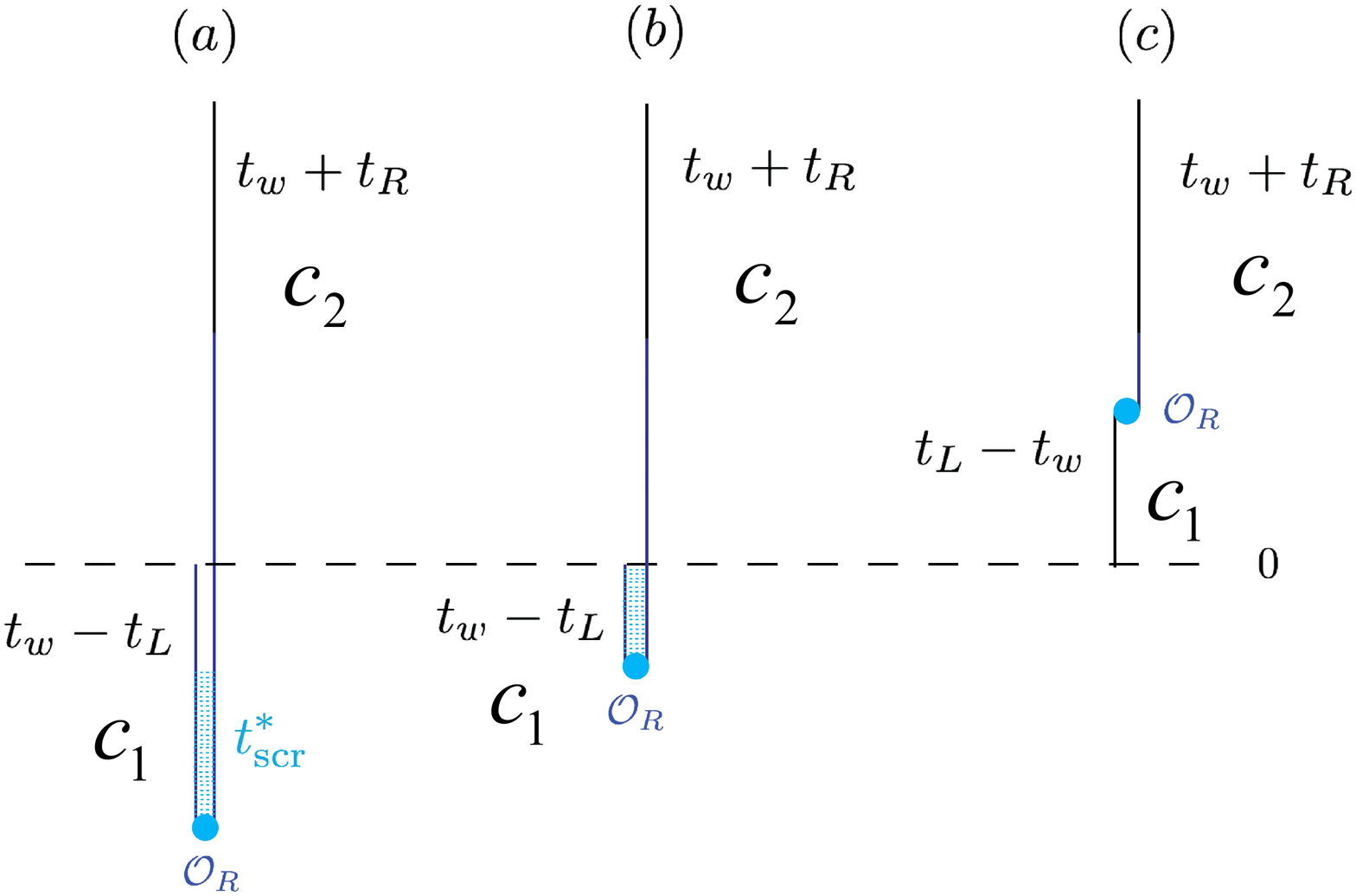}
\caption{A representation of the insertion of a  perturbed operator $\math{O}_R$ at the time $-t_w$ for the TFD state.}\label{swc}
\end{figure}
\subsection{Circuit analogy}\label{circuit}
In this subsection, we would like to investigate the connection between the behaviours of our holographic results and the switchback effect of the circuit model. As discussed in Ref.\cite{F1}, evolving the perturbed state independently in the left and right times yield the expression
\ba\label{SBP}
\bra{TFD(t_L,t_R)}_\text{pert}=U_R(t_R+t_w)\math{O}_RU_R(t_L-t_w)\bra{TFD}\,,\nn
\ea
where the perturbed operator $\math{O}_R$ is a localized simple operator. $U_R(t)\math{O}_R U_R(-t)=I$ with the identity operator $I$ when $t<t^*_\text{scr}$. This feature is connected to the switchback effect and can provide a deeper explanation of our holographic results.

We denote the rate of the complexity to $c_1$ before the operator $\math{O}_R$ is inserted and $c_2$ after it.  Under the limit of light shock, we have $c_1\approx c_2\approx c$.

First of all, we consider the case $t_w<t^*_\text{scr}$. When $t_L<t_w$, the process in \eq{SBP} can be illustrated in (b) of \fig{swc}. In this situation, the switchback effect produces a cancellation for the process below the dashed line. Therefore, the complexity is given by
\ba\label{cpert1}
\math{C}_\text{pert}\approx 2 c\, t\,,
\ea
where we set $t_L=t_R=t/2$. One can note that this complexity is exactly the result of the eternal case where the cancellation is always valid for the process below the dashed line. When $t_L>t_w$, the process can be illustrated by (c) in \fig{swc}. We can see that there is no opportunity for the switchback effect. Hence, the complexity is also the result of the eternal case which can be described by \eq{cpert1}. As a summary, we find that when $t_w<t^*_\text{scr}$, by virtue of the switch back effect, the complexity is same as that of the unperturbed state. This behavior is in agreement with our holographic result represented by the left panel of \fig{LdCdttw}.

Then, we consider the case $t_w>t^*_\text{scr}$. When $t_L-t_w>-t^*_\text{scr}$, the complexity shares the same result with the case $t_w<t^*_\text{scr}$. When $t_w-t_L>t^*_\text{scr}$, the process can be illustrated by (a) in \fig{swc}. In this case, the two time-evolution operators cancel out only during the scrambling time. Therefore, the complexity can be written as
\ba
\math{C}_\text{pert}\approx 2c\,(t_w-t^*_\text{scr})\,.
\ea
This result shows that the growth rate is very close to zero in the region $t<2(t_w-t^*_\text{scr})$. This feature is in agreement with our holographic result as shown in right panel of \fig{dCdttw}.

Next, we consider the complexity of formation. By setting $t=0$ an using the above equations, one can obtain
\ba
\frac{d\D \math{C}_\text{pert}}{d t_w}=2c\math{H}(t_w-t^*_\text{scr})\,.
\ea
Again, this formula also matchs the our holographic case as illustrated in left panel of \fig{dC} in which when $t<t^*_\text{scr}$, the rate of the complexity of formation is close to zero, and when $t>t^*_\text{scr}$, it remains constant.

\section{Conclusion and discussion}\label{sec6}
The action of AdS black hole within the WDW patch has been related to the quantum complexity of a holographic state. Following the procedure in \cite{A1}, we calculated the action growth rate of the charged AdS-Vaidya black hole in $(d+1)$-dimensional Einstein-Maxwell gravity. We first introduced a charged AdS-Vaidya geometry which is source by the collapse of an uncharged thin shell of null fluid. And this thin shell generates a shape transition from a black hole with total mass $M_1$ and charge $Q$ to another one with mass $M_2$ and the same charge $Q$.

Using the approach proposed by Lehner $et\, al.$ \cite{A5}, we studied the complexity of the formation and discussed its small and large time behaviors in Sec.\ref{complexiyofformation}. We found that the slope of the complexity of formation shares the similar behaviors with the uncharged case. Meanwhile, these results are also in agreement with the switchback effect.
After that, the growth rate of the complexity was evaluated in Sec.\ref{evocom}. By comparing it to the uncharged case, we found that the behaviors for the charged cases can smoothly approach that of the neutral cases. Furthermore, we also found that when $t_w<t^*_\text{scr}$, the action growth rate is the same as the unperturbed case, and when $t_w>t^*_\text{scr}$, it shares the similar behaviors with the heavy shock wave case. And these behaviors can be explained by the switchback effect. In addition, we show that the late time growth rate is given by the average value of the two RN rate without shockwave, which is consistent with the uncharged case.
In Sec.\ref{without}, we investigated the early and late time behaviors of the complexity without the counterterm. We demonstrated that, in order to obtain an expected property of the complexity, it is also necessary to introduce the counterterm on the null boundaries for the charged Vaidya black hole. Finally, by analysing the circuit model, we showed our results our holographic results are in agreement with that of the circuit model.

In this paper, we only considered the CA conjecture in charged RN black hole sourced by the collapse of an uncharged thin shell of null fluid. It would also be interesting to further investigate the CV conjecture in the charged Vaidya black hole. As discussed in the uncharge case \cite{A1}, the CV conjecture also shares the similar results with the CA conjecture, such as the late time behaviors and the switch back effect. Therefore, we have good reason to believe that the CV conjecture have same behaviors with the CA conjecture in the charged Vaidya black hole, such as the late time action growth rate can also be expressed as the sum of the average value of the two RN rate without shockwave. In addition, it would be interesting to investigate the charged Vaidya black hole with a charged shock wave, in which we might possible to study the one-side charged Vaidya spacetimes which formed by the collapse of an charged spherically symmetric shell to the AdS vacuum spacetime, and consider the process from the finite temperature black hole to extremal black hole.

\section{Acknowledgments}
 This research was supported by NSFC Grants No. 11775022 and 11375026. The author is grateful to the anonymous referees for their useful comments which have significantly improved the quality of our paper.

\end{document}